\documentclass[aps,prl,superscriptaddress,twocolumn,showpacs]{revtex4}

\usepackage[dvips]{graphicx}
\begin{document}


\title{The Spontaneous Generation of Magnetic Moment of Beryllium Dimer on Graphene}


\author{Xiang He}
\affiliation{Eco-Materials and Renewable Energy Research Center, Department of Physics, National Laboratory of Solid State Microstructures, Nanjing University, China.}
\author{Zhao-Xu Chen}
\affiliation{Institute of Theoretical and Computational Chemistry, Key Lab of Mesoscopic Chemistry of Ministry of Education, Department of Chemistry, Nanjing University, China.}
\author{Zhaosheng Li}
\affiliation{Eco-Materials and Renewable Energy Research Center, Department of Physics, National Laboratory of Solid State Microstructures, Nanjing University, China.}
\author{Zhigang Zou}
\email[]{zgzou@nju.edu.cn}
\affiliation{Eco-Materials and Renewable Energy Research Center, Department of Physics, National Laboratory of Solid State Microstructures, Nanjing University, China.}


\date{\today}

\begin{abstract}
Graphene has various potential applications in electronics and scientists are seeking to introduce magnetic properties on graphene. Here we report our theoretical findings that a local magnetic moment as large as 1 $\mu_B$ can be generated from the adsorption of the diamagnetic beryllium dimer on perfect diamagnetic graphene. Unexpectedly the adsorption system is remarkably stable as evidenced by the binding energy of 1.04 eV, indicating that the formation of beryllium dimer is energetically much favorable. Analyses reveal that the magnetic moment and stability is originated from the spontaneous transfer of one electron from the anti-bonding orbital of the dimer to graphene.
\end{abstract}


\maketitle

Graphene has become a promising material for future electronics. The special electronic and magnetic properties of graphene are related to its low-dimensional carbon structures and have aroused enormous interest.\cite{geim:nmat,geim:science,castro:rmp} Since the pure graphene is diamagnetic, introducing magnetic properties to graphene-related systems has attracted theoretical concern.\cite{mcclure:pr,brey:prl,uchoa:prl} By doping magnetic transition-metal elements on perfect graphene or graphene vacancies one can introduce magnetic moment into graphene system.\cite{krasheninnikov:prl,santos:prb,sevincli:prb} Adsorbing small covalent molecules on perfect graphene can also change the total magnetic moment by charge transfer between adsorbates and graphene.\cite{leenaerts:apl,leenaerts:prb,wehling:nanolett} It is notable that these covalent molecules stay away from graphene layer by 3 \AA\ at least and therefore the binding energies are small. Typically, the transferred charges are less than 0.1 e$^-$. Thus, the induced change of magnetic moment is expectedly small.\cite{leenaerts:apl,leenaerts:prb} It is highly desired to find an adsorbate that can adsorb on graphene firmly and induce a large charge transfer or a remarkable magnetic moment. With density functional theory calculations, we found a large local magnetic moment can be produced spontaneously by adsorbing diamagnetic beryllium dimer on perfect diamagnetic graphene. Here we report this interesting finding.

All the calculations were performed with the plane wave based VASP code\cite{kresse:prb48,kresse:prb50,kresse:cms,kresse:prb54} using the projector-augmented wave method\cite{blochl:prb,kresse:prb59} and Perdew-Burke-Ernzernhof generalized gradient approximation\cite{perdew:prl}. The supercell is composed of $4\times4$ unit cells with a distance of 15 \AA\ between the adjacent graphene layers. For the integration over the Brillouin zone, we combined $5\times5\times1$ Monkhorst-Pack grids\cite{monkhorst:prb} with a generalized Methfessel-Paxton smearing technique\cite{methfessel:prb} to optimize adsorption geometries. Charge transfers were calculated based on the Bader charge analysis.\cite{bader:aqc,tang:jpm} For accurate charge and energy calculations a more fine $11\times11\times1$ Monkhorst-Pack grids with tetrahedron method with Bl\"{o}chl corrections were used. An energy cutoff of 600 eV was used throughout the calculations.

Beryllium dimer is the simplest metal dimer in addition to lithium dimer. It has eight electrons totally and the electronic configuration of the ground state is $(1\sigma)^2(1\sigma^*)^2(2\sigma)^2(2\sigma^*)^2$. The ground state of Be$_2$ is singlet since its electronic configuration is a type of closed shell. In old-fashioned text book, Be$_2$ molecule cannot exist in reality because the full-filled anti-bonding $2\sigma^*$ orbital counteracts the full-filled bonding $2\sigma$ orbital energetically. Since 1980s, theoretical studies gradually indicate the existence of the stable Be$_2$ with a short bond length by mixing excited determinants.\cite{roeggen:ijqc60,roeggen:ijqc101} Recent experimental work determined potential energy curve of Be$_2$ and found it looks like a normal covalent molecule of weak bond strength with a bond distance of 2.45 \AA.\cite{merritt:science} Our calculated bond length of beryllium dimer is 2.43 \AA, in nice agreement with the experimental value.

\begin{figure}[htbp]
\includegraphics[scale=0.3]{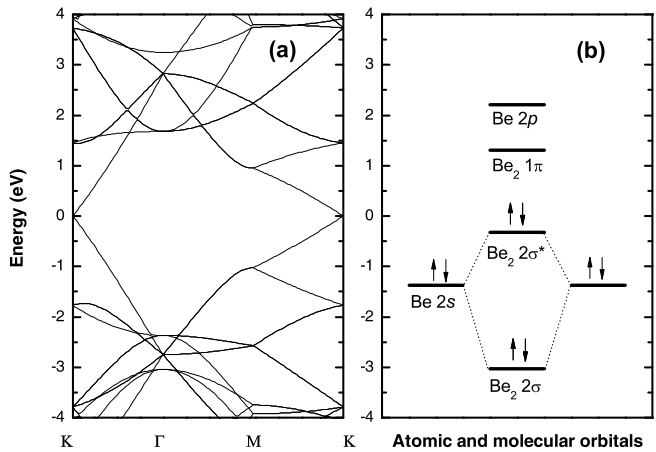}%
\caption{\label{F1}(a) The band structure of graphene. (b) The energy levels of Be 2s, 2p atomic orbitals and Be$_2\ 2\sigma,\ 2\sigma^*,\ 1\pi$ molecular orbitals.}
\end{figure}

Fig. \ref{F1}(a) shows the band structure of pure graphene. As comparison, the relative energy level of isolated Be 2s, 2p atomic orbitals and Be$_2$ $2\sigma,\ 2\sigma^*,\ 1\pi$ molecular orbitals are also presented in Fig. \ref{F1}(b). The full-filled $2\sigma^*$ orbital is 0.33 eV lower than the Dirac point of graphene while the empty $1\pi$ orbital is 1.30 eV higher than the Dirac point. Such energy level ordering should prevent charge transferring from Be$_2$ to graphene or vice versa. In fact, in the physisorption processes of diamagnetic covalent molecules, such as H$_2$O, NH$_3$ and CO, the highest occupied molecular orbital (HOMO) and lowest occupied molecular orbital (LUMO) of the adsorption system are separated by the Fermi level so the total charge transfer is small.\cite{leenaerts:prb}

\begin{figure}[htbp]
\includegraphics[scale=0.4]{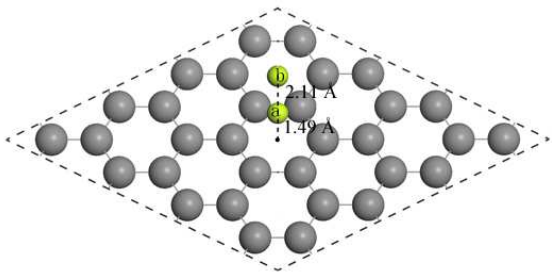}%
\caption{\label{F2}The stable adsorption of beryllium dimer on the ($4\times4$) cell of graphene. The dimer is perpendicular to the graphene and adsorbs on the hollow site. Be$_a$ atom is below Be$_b$ atom. Labeling of atoms: C, gray spheres; Be, yellow spheres.}
\end{figure}

Now we study beryllium dimer adsorbed on graphene. Our calculation revealed that though one single beryllium atom can hardly adsorb on graphene, the beryllium dimer does adsorb on graphene. We have investigated several adsorption configurations. The most stable adsorption geometry is shown in Fig. \ref{F2}. The Be-Be bond length on graphene is shortened by 0.32 \AA\ compared with that in gas phase. The perpendicular distance to graphene layer is 1.49 \AA\ (Fig. \ref{F2}) which is significantly shorter than those for other molecules doped on graphene.\cite{leenaerts:apl,leenaerts:prb,wehling:nanolett,zanella:prb} The binding energy (BE) of the dimer is unexpectedly as high as 1.04 eV. ($BE=E_{Be_{2}}+E_{graphene}-E_{Be_{2}-graphene}$) Apart from large binding energy, it is also amazing that when the singlet diamagnetic beryllium dimer adsorbs on diamagnetic graphene, the Be$_2$-graphene adsorption system acquires magnetic moment of 1 $\mu_B$ per supercell. The emergence of magnetic moment usually indicates selective charge transfer of spin channel, which is contrary to our previous prediction based on the energy spectra of isolated beryllium dimer and graphene.

\begin{figure}[htbp]
\includegraphics[scale=0.3]{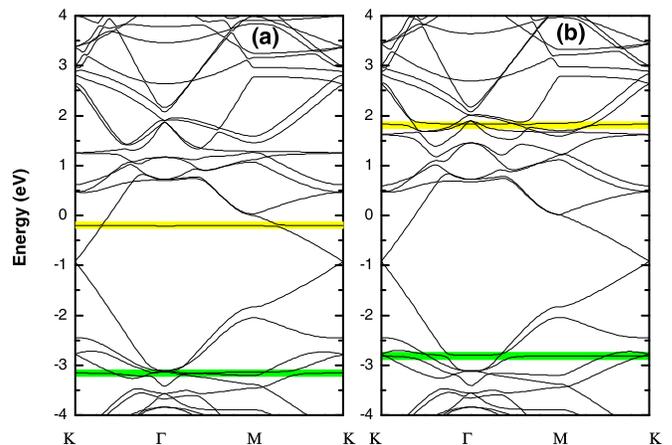}
\caption{\label{F3}The band structure of Be$_2$-graphene system. (a) The spin-up component. (b) The spin-down component. The $2\sigma$ and $2\sigma^*$ orbitals of Be$_2$ are marked in green and yellow, respectively.}
\end{figure}

\begin{figure}[htbp]
\includegraphics[scale=0.3]{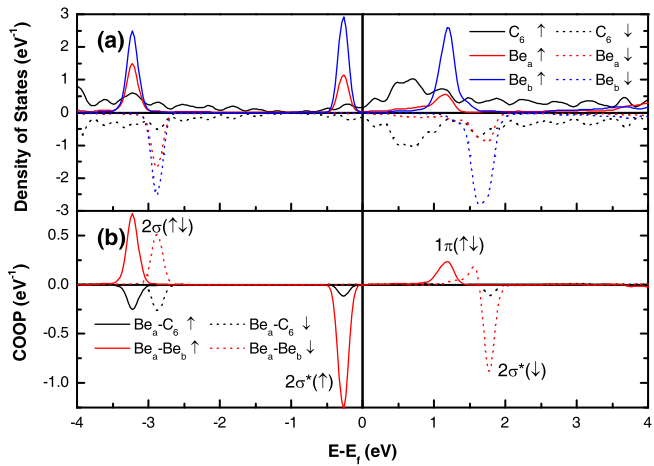}
\caption{\label{F4}Comparison of local DOS and COOP. (a) Local DOS of two beryllium atoms and six carbon atoms around the adsorption sites. (b) COOP in red is of two beryllium atoms. The black one is the total COOP between Be$_a$ (see the caption of Fig. \ref{F2}) and six carbon atoms around the adsorption site.}
\end{figure}

The band structure of Be$_2$-graphene system is illustrated in Fig. \ref{F3}. The upshift of the Fermi level relative to the Dirac point results from the charge transfer from Be$_2$ molecule to graphene. One can also find the splitting of $2\sigma$ and $2\sigma^*$ spin orbitals of Be$_2$ (see Fig. \ref{F3}). The $2\sigma$ spin orbitals are still lower than the Dirac point compared with the one in gas phase, but the $2\sigma^*$ spin orbitals are pushed higher than the Dirac point. Fig. \ref{F3}(a) shows the spin-up component of $2\sigma^*$ ortital is below the Fermi level, while the spin-down component of $2\sigma^*$ ortital in Fig. \ref{F3}(b) is higher than the Fermi level. Especially, the spin-down component of $2\sigma^*$ orbital is much higher than the Dirac point. Thus it is the electron of spin-down component on $2\sigma^*$ orbital which is transferred to the conduction band of graphene spontaneously. In Fig. \ref{F4}(a) we plot the total density of states (DOS) of two beryllium atoms and the total DOS of six carbon atoms around the hollow site. It is not difficult to verify that the split of $2\sigma$ spin ortitals is about 0.35 eV, the $2\sigma^*$ orbital of spin-up component is lower than the Fermi level by 0.27 eV and the $2\sigma^*$ orbital of spin-down component is higher than the Fermi level by 1.77 eV. Using Bader charge analysis we found three valance electrons stay on Be$_2$ and the total magnetic moment of 1 $\mu_B$ localizes on Be$_2$ in the Be$_2$-graphene system which means one electron of spin-down component is transferred to graphene totally.

It is instructive to compare the adsorption of paramagnetic NO$_2$ and diamagnetic Be$_2$ on graphene. The mechanisms of the charge transfer are different for the two adsorbates. In the case of the adsorption of the paramagnetic NO$_2$ on graphene, it was argued that since the empty partially occupied molecular orbital (POMO) of adsorbed NO$_2$ is below the Dirac point in the adsorbed configuration, the charge transfer can reach one electron in the dilute limit.\cite{leenaerts:apl,wehling:nanolett,schedin:nmat} In the actual finite theoretical calculation, the empty POMO is crossed by the Fermi level so just fractional charge is transferred to the POMO from graphene and the total magnetic moment decreases.\cite{leenaerts:apl,leenaerts:prb} In the Be$_2$-graphene system, the Fermi level locates between the two distinctly split $2\sigma^*$ spin orbitals of Be$_2$ in the adsorbed configuration. Hence, one electron is transferred from the spin orbital above the Fermi level to graphene and a magnetic moment of 1 $\mu_B$ is produced spontaneously.

Large binding energy of doping adsorbates on graphene usually involves the strong interaction between adsorbate atoms and carbon atoms of graphene, such as the adsorption of transition-metal elements.\cite{santos:prb} To check whether this is the case in the Be$_2$-graphene system, we also plot the total crystal orbital occupation populations (COOP) of adsorbed Be$_2$ and the one between Be$_2$ and six carbon atoms around the adsorption site in Fig. \ref{F4}(b). COOP is a more illustrative scheme to reflect the nature and strength of bonding or anti-bonding interaction between two bonding atoms than DOS.\cite{hughbanks:jacs} The COOP plots are characterized with $2\sigma$ and $2\sigma^*$ of Be$_2$ orbitals. The $2\sigma^*$ spin orbitals are split by 2.04 eV which coincides with the DOS analysis. From Fig. \ref{F4}(b) we can infer that the interaction of Be$_2$ $2\sigma^*$ orbitals with graphene carbons is anti-bonding and negligible. The adsorption of beryllium dimer on graphene is a kind of physisorption essentially. But why the binding energy calculated is much larger than that of other small covalent molecule doping on graphene where the binding energies is an order of meV? This is because the electron transferred from Be$_2$ to graphene is from the anti-bonding $2\sigma^*$ orbital. The removal of one electron from the anti-bonding orbital stabilizes Be$_2$ and yields large binding energy. The shortening of the Be-Be bond length by 0.32 \AA\ provides a strong support for this point of view.

By far, the study presented here only considers the ferromagnetic configuration of the Be$_2$-graphene system. In order to check the magnetic state of the adsorption structure, we doubled the geometry in one direction and set the initial magnetic moments of beryllium to be antiferromagnetic. Calculation results showed the energies of both configurations are nearly equal, which means the size of supercell in our calculations is large enough to omit the magnetic couplings between Be$_2$ of each supercell. In addition, the spin-unpolarized configuration was also considered and found unfavorable energetically.

In summary, we found beryllium dimer is favorable to form on graphene. The physisorption of diamagnetic beryllium dimer on pure graphene induces a charge transfer of one electron and generates a magnetic moment of 1 $\mu_B$. The magnitude of charge transfer and magnetic moment generated by doping beryllium dimer are larger than by doping those reported molecules on graphene. Our study demonstrates that even without transition-metal adatoms or defective graphene a large magnetic moment can be generated spontaneously from two diamagnetic matters. This opens up the possibility of introducing stable and observable magnetic properties, such as magnetic domain and magnetic order, in graphene by molecular doping directly.

This work is supported by The National Basic Research Program of China (973 Program, 2007CB613301, 2007CB613305). The national natural science foundation of China No.20973090 is gratefully acknowledged.

\bibliography{Be2G}

\end{document}